\documentclass[prd,nofootinbib]{revtex4}
\usepackage{epsfig}
\usepackage{graphicx}
\usepackage{epstopdf}
\newcommand{\ud}{\mathrm{d}}

\begin{document}

\title{Accurate calculations of the WIMP halo around the Sun and prospects for its gamma-ray detection}

\author{Sofia Sivertsson}
\email{sofiasi@kth.se}
\affiliation{Department of Theoretical Physics, Royal Institute of Technology (KTH), AlbaNova University Center, 106 91 Stockholm, Sweden}
\affiliation{The Oskar Klein Centre for Cosmoparticle Physics, Department of Physics, Stockholm University, AlbaNova University Center, 106 91 Stockholm, Sweden}

\author{Joakim Edsj{\"o}}
\email{edsjo@fysik.su.se}
\affiliation{The Oskar Klein Centre for Cosmoparticle Physics, Department of Physics, Stockholm University, AlbaNova University Center, 106 91 Stockholm, Sweden}

\date{\today}

\begin{abstract}
Galactic weakly interacting massive
particles (WIMPs) may scatter off solar nuclei to orbits gravitationally bound to the Sun.
Once bound, the WIMPs continue to lose energy by repeated scatters in
the Sun, eventually leading to complete entrapment in the solar interior.  
While the density of the bound population
is highest at the center of the Sun, the only observable signature of
WIMP annihilations inside the Sun is neutrinos.  It has been previously
suggested that although the density of WIMPs just outside the Sun is
lower than deep inside, gamma rays from WIMP annihilation just outside
the surface of the Sun, in the so called WIMP halo around the Sun, may be more easily detected.
We here revisit this problem using detailed Monte Carlo simulations and detailed composition and structure information about the Sun to estimate the size of the gamma-ray flux. Compared to earlier simpler estimates, we find that the gamma-ray flux from WIMP annihilations in the solar WIMP halo would be negligible; no current or planned detectors would be able to detect this flux.
\end{abstract}

\maketitle

\section{Introduction}
The observational evidence for dark matter is overwhelming (see e.g.\ \cite{Bergstrom:2000pn}), but in spite of that, direct evidence for dark matter particles has not yet been found. One of the most popular classes of dark matter models is the Weakly Interacting Massive Particle (WIMP) model. WIMPs are heavy particles with weak interaction strengths, and are popular because they are naturally thermally produced at right abundances in the early Universe (see e.g.\ \cite{Bergstrom:2009ib}). There are many ways to search for WIMPs, e.g.\ via direct detection, or via indirect detections, like neutrinos from the Sun \cite{1985ApJ...296..679P,1985PhRvL..55..257S} or the Earth \cite{1986PhLB..167..295F,1986PhRvD..33.2079K,1986PhRvD..34.2206G}, gamma-rays from annihilations in the Galactic halo or charged cosmic-rays from annihilations in the Galactic halo (for a review, see \cite{Bergstrom:2000pn}). The neutrino signal from the Sun arises from WIMPs scattering in the Sun, eventually sinking to the core and annihilating, producing many standard model particles. Of these, only neutrinos escape the Sun and the neutrinos can be searched for with neutrino telescopes like e.g.\ IceCube \cite{2004APh....20..507A}. 

We will here focus on another idea, proposed by Strausz \cite{Strausz}. The idea is that while WIMPs are being captured by the Sun, they will move on bound orbits outside the Sun, making up a halo of WIMPs around it. Pair-wise annihilations of WIMPs in this halo could occur and (if the density would be high enough), produce a detectable flux of e.g.\ gamma-rays. This type of gamma-ray flux has been searched for by the Milagro \cite{milagro} detector.

Strausz came to the conclusion that this gamma-ray signal would be measurable or that if no such signal is seen, one would be able to constrain parameters in the WIMP model. However, Strausz' calculation imposes many approximations and in a later preprint by Hooper \cite{Hooper}, it was found that the flux of gamma-rays is much smaller than Strausz' estimates and a detection of this gamma-ray signal would require unrealistically large telescope areas. In both these calculations many simplifying approximations are made; for example radial orbits are assumed, the WIMPs real orbits are not followed with realistic solar models. Another calculation was performed by Fleysher \cite{Fleysher:2003iya}, finding even higher gamma-ray fluxes than in Strausz' calculation. However, many simplifying assumptions were made in all of these calculations, making the results unreliable.
%However, one can easily show that that calculation is incorrect (see section \ref{sec:disc_prev} for a discussion about this), and gives unreasonably large fluxes.

To clarify the discrepancies between these earlier estimates, we have here performed a much more thorough calculation of the WIMP density around the Sun via detailed Monte Carlo simulations. We let WIMPs from the Galactic halo scatter in the Sun and follow the WIMPs on their orbits (taking full account of the non-Keplerian nature of the orbits inside the Sun), letting them interact repeatedly with solar nuclei until they are fully trapped inside the Sun. We calculate how much each WIMP we simulate contributes to the WIMP density around the Sun. By summing up the contributions from the entire WIMP population, we can estimate the total WIMP density around the Sun. We finally calculate the gamma-ray signal from WIMP annihilations in this halo, which can be compared with the Milagro searches for a gamma-ray signal of this kind \cite{milagro}. We will also calculate predicted fluxes for future gamma-ray detectors and compare with expected backgrounds.

%%%%%%%%%%
\section{The WIMP capture process}

\begin{figure}
\centerline{\epsfig{file=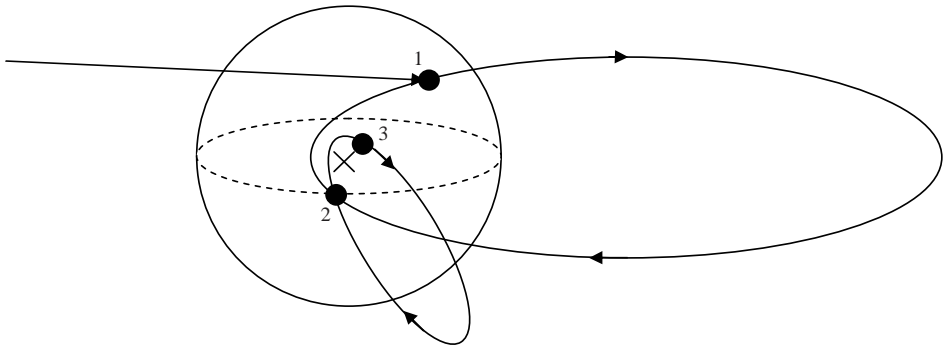,width=0.7\textwidth}}
\caption{An example of a WIMP being captured by the Sun. A WIMP from the Milky Way halo population scatters at point 1 as it passes through the Sun. If enough energy is lost the WIMP ends up on a bound orbit which it travels in, typically for many solar passages, until it scatters again in point 2 to an orbit with lower energy. In this example the WIMP has, after the scatter in point 3, lost enough energy to become completely trapped inside the Sun and is no longer of any interest to us. Note that in the figure the orbits are entirely elliptical, whereas in reality (and in our simulations), they are slightly non-elliptical inside the Sun.}
\label{fig:WIMPstart}
\end{figure}

As Milky Way WIMPs pass through the Sun, some will scatter off solar nuclei and some of those will lose enough energy to become gravitationally bound to the Sun's potential well. The bound WIMPs will orbit the centre of the Sun and eventually scatter again in the Sun, losing energy in each scatter. Eventually enough energy is lost for the WIMP to end up on an orbit which lies completely inside the Sun (at which point we will called it completely trapped). The WIMP capture process is illustrated in Fig.~\ref{fig:WIMPstart}. 

Due to the smallness of the scatter cross section, a bound WIMP typically survives many passages in the Sun before it scatters again. Also, as the WIMP is typically much heavier than the nuclei it scatters off the energy loss in each scatter tend to be quite small. Because of this the WIMPs often have to scatter many times before enough energy is lost for the WIMP to become completely trapped inside the Sun. In this way a captured WIMP can contribute to the WIMP density around the Sun for a long time even though a small energy loss in the first scatter can imply that the first orbit stretches very far out from the Sun. The effects of small energy losses in each scatter is most pronounced if hydrogen is an important target for WIMP scattering in the Sun.

To get good accuracy of our predictions, we follow the WIMPs through the Sun on their actual orbits, taking the actual solar environment (as a function of radius), and details of the scattering process into account. 

All these intermediate orbits contribute to the WIMP density around the Sun, building up a WIMP halo around it (we will refer to this population of WIMPs as the solar WIMP halo, or just WIMP halo, not to be confused with the galactic WIMP halo, or galactic halo for short). WIMP annihilations within this halo give rise to high-energy gamma-rays, where the dominant background is cosmic-ray interactions with the solar chromosphere \cite{1991ApJ...382..652S} and up-scattering of solar photons via inverse-Compton scattering from cosmic-ray electrons \cite{Orlando:2006zs,Moskalenko:2006ta,2007ApJ...664L.143M}.
Also, the Sun is opaque to gamma-rays and hence shields against the diffuse gamma-ray background.

%%%%%%%%%%
\section{Build-up of the solar WIMP halo}

\subsection{The WIMPs' first scatter in the Sun}
Following the notation of Gould \cite{Gould}, the total number of WIMPs that scatter in a shell of a given radius in the Sun per unit time and velocity is given by
\begin{equation}
\frac{\ud^2 \Gamma}{\ud r \ud u} = 
4\pi r^2\frac{\sigma\rho(r)}{m}w^2\frac{f(u)}{u} \label{vel_int},
\end{equation}
where $f(u)$ is the velocity distribution of the WIMPs in the Milky Way halo, in the Sun's reference system and infinitely far away, $m$ is the mass of the nucleus in the Sun which the WIMP scatters off, $\sigma$ is the scatter cross section on that nucleus, $u$ is the WIMP velocity at infinity, $v$ is the escape velocity at the shell, $w$ is the velocity of the WIMP in the shell, $r$ is the radius of the shell and $\rho(r)$ is the density of the element with nucleus mass $m$ at radius $r$. We will later denote the WIMP mass with $M$. Eq.~(\ref{vel_int}) should be summed over all elements in the Sun. We use the solar model of \cite{Bahcall:2004pz} with heavy element abundances from \cite{1998SSRv...85..161G} as implemented in DarkSUSY \cite{darksusy}. In total we include the 16 most important elements in the Sun in our calculation.

Assuming the velocities of the WIMPs in the Milky Way halo to be Maxwell-Boltzmann distributed, the velocity distribution in the Sun's reference frame is \cite{Gould}
\begin{equation}
\frac{f(u)}{u}\,\ud u=n_\mathrm W\sqrt{\frac{6}{\pi}}\frac{1}{v_\odot\bar{v}}\exp\left(-\frac{3v_\odot^2}{2\bar{v}^2}\right)
\exp\left(-\frac{3u^2}{2\bar{v}^2}\right)\sinh\left(\frac{3uv_\odot}{\bar{v}^2}\right)\,\ud u \label{vel-distr},
\end{equation}
where $v_\odot$ is the Sun's velocity orbiting the Milky Way centre and $\bar v$ is the WIMP three-dimensional velocity dispersion in the Galactic halo, here taken to be $\bar v=270$ km/s. $n_\mathrm W=\rho_\mathrm W/M$ is the number density of WIMPs in our region of the Milky Way halo, where we take  $\rho_\mathrm W=0.3$~GeV~cm$^{-3}$. 

For further use the WIMPs' energy distribution will be more useful than the velocity distribution; Eq.~(\ref{vel-distr}) can be rewritten as
\begin{equation}
\frac{f_{\mathcal E}(\mathcal E)}{\mathcal E}\,d\mathcal E = n_\mathrm W\sqrt{\frac{3}{\pi}}\frac{1}{v_\odot\bar{v}}\exp\left(-\frac{3v_\odot^2}{2\bar{v}^2}\right)
\exp\left(-\frac{3\mathcal E}{\bar{v}^2}\right)\sinh\left(\frac{3v_\odot}{\bar{v}^2}\sqrt{2\mathcal E}\right)\frac{1}{\sqrt{\mathcal E}}\,d\mathcal E,
\end{equation}
where $\mathcal E$ is the reduced energy, defined as the WIMP's energy (kinetic plus potential) divided by the WIMP mass (for a captured WIMP $\mathcal E < 0$). Analogously $\mathcal E_\mathrm k$ is the reduced kinetic energy and $\mathcal E_\mathrm p$ is the reduced potential energy at some radius.

Rewriting Eq.~(\ref{vel_int}) now yields the number of WIMPs which scatter in a shell in the Sun per unit time and reduced energy 
\footnote{ Note that
$$
\lim_{u \to 0}\frac{f(u)}{u}=0 \mbox{ \ but \ } \lim_{\mathcal E \to 0} \frac{f_{\mathcal E}(\mathcal E)}{\mathcal E}\neq 0 %\label{footnote}
$$
and hence the number of scatters per unit time and velocity vanishes as $u\to 0$ while the number of scatters per unit time end energy does not vanish as $\mathcal E\to 0$. }
\begin{equation}
\frac{\ud^2\Gamma}{\ud\mathcal E\, \ud r}=8\pi r^2\frac{\sigma\rho(r)}{m}(\mathcal E-\mathcal E_\mathrm p) \frac{f_{\mathcal E}(\mathcal E)}{\mathcal E} \label{energy_int}
\end{equation}
expressed using the reduced energy the WIMP has before it scatters.

%%%%%
\subsubsection{The WIMPs' energy distribution after their first scatter}
When a WIMP interacts with a nucleus it scatters from the reduced kinetic energy before the scatter, $\mathcal E_\mathrm k$, to the reduced kinetic energy after the scatter, $\tilde \mathcal E_\mathrm k$, which, as for general elastic scatter, fulfils the relation
\begin{equation}
\left(\frac{M-m}{M+m}\right)^2\mathcal E_\mathrm k\leq \tilde \mathcal E_\mathrm k \leq \mathcal E_\mathrm k \label{ineq-scatter}
\end{equation}
with all $\tilde \mathcal E_\mathrm k$ in this interval being equally probable. Note that it is the energy loss and not the loss in velocity that is uniformly distributed.

This means that a WIMP with the reduced energy $\tilde \mathcal E_\mathrm k$ can only have scattered from reduced energies fulfilling
\begin{equation}
\tilde \mathcal E_\mathrm k\leq \mathcal E_\mathrm k\leq \left(\frac{M+m}{M-m}\right)^2\tilde \mathcal E_\mathrm k,
\end{equation}
which written in terms of total reduced energy $\mathcal E=\mathcal E_\mathrm k+\mathcal E_\mathrm p$ (and $\tilde \mathcal E=\tilde \mathcal E_\mathrm k+\tilde \mathcal E_\mathrm p$, with $\tilde \mathcal E_{\rm p} = \mathcal E_{\rm p}$) becomes
\begin{equation}
\tilde \mathcal E\leq \mathcal E \leq\left(\frac{M+m}{M-m}\right)^2\tilde \mathcal E-\frac{4Mm}{(M-m)^2}\mathcal E_\mathrm p. \label{limits}
\end{equation}
Note that the reduced energies are greater than zero for particles not bound to the solar system, hence $\mathcal E \geq 0$ for the WIMPs in the Milky Way halo, whereas $\tilde \mathcal E$ can be both negative or positive, depending on the energy lost in the scatter.

%Note that $\mathcal E\geq 0$ since the particles in the Milky Way halo are not bound to the Sun, while the sign of the reduced energy after the scatter, $\tilde \mathcal E$, is not determined.

The number of WIMPs which scatter to a certain reduced energy, $\tilde \mathcal E$, in a given shell in the Sun, per unit time is then 
\begin{eqnarray}
\frac{\ud^2\Gamma}{\ud\tilde\mathcal E\,\ud r} &=& 2\pi r^2\frac{\rho(r)}{m}\frac{(m+M)^2}{Mm}
\int_{\tilde{\mathcal E}}^{u.l.}\sigma\frac{f_{\mathcal E}(\mathcal E)}{\mathcal E}\theta(\mathcal E)\,\ud\mathcal E \label{integral1}, \\
\mbox{where \ \ \ }
u.l. &=& \left(\frac{M+m}{M-m}\right)^2\tilde \mathcal E-\frac{4Mm}{(M-m)^2}\mathcal E_{\mathrm p} \nonumber
\end{eqnarray}
which is the integral of (\ref{energy_int}) with integration limits given by (\ref{limits}). Here $\theta$ is the Heaviside function, which is required since $\tilde\mathcal E$ can be negative while $\mathcal E$ is always positive.

The above expression has been normalised by multiplying by the reciprocal of the interval length in (\ref{ineq-scatter})
\begin{equation}
\frac{1}{\mathcal E_\mathrm k}\frac{(M+m)^2}{4Mm} \label{expr-probability}.
\end{equation}

%%%%%%%%%%
\subsubsection{Form factor suppression of the scattering cross section}

For scatter off nuclei larger than the proton the cross section depends on the energy loss in the scatter. For scatters with low momentum transfer the wave functions of the different nucleons in the nuclei may be viewed as coherent, making the cross section increase as the square of the number of nucleons in the nucleus, $A^2$. For higher momentum transfer this coherence breaks down. This suppression is taken into account by introducing an exponential form factor, suppressing high-energy transfers, i.e.\ high $\Delta\mathcal E$ \cite{Gould}
\begin{equation}
\exp\left( -\frac{\Delta\mathcal E}{\mathcal E_0}\right) \label{form_factor}
\end{equation}
with
\begin{equation}
\mathcal E_0\equiv\frac{3\hbar^2}{2mMR^2}\mbox{ \ and \ }
R\sim \left[0.91\left(\frac{m}{GeV}\right)^{1/3}+0.3\right]\times 10^{-15} \mbox{ \ m},
\end{equation}
where $R$ is an estimate of the radius of the nucleus \cite{Eder}. 

Introducing a form factor, Eq.~(\ref{integral1}) can be written as
\begin{eqnarray}
\frac{\ud^2\Gamma}{\ud\tilde\mathcal E\,\ud r}&=&2\pi r^2\frac{\rho(r)}{m}\tilde{\sigma}\frac{(m+M)^2}{Mm}
\int_{\tilde\mathcal E}^{u.l.}\frac{f_{\mathcal E}(\mathcal E)}{\mathcal E}\exp\left(\frac{\tilde \mathcal E- \mathcal E}{\mathcal E_0}\right)\theta(\mathcal E)\ud\mathcal E, \mbox{ \ \ }\label{re-expr-capture}\\
\mbox{ \ where \ }
u.l. &=& \left(\frac{M+m}{M-m}\right)^2\tilde \mathcal E-\frac{4Mm}{(M-m)^2}\mathcal E_\mathrm p \nonumber
\end{eqnarray}
where $\tilde{\sigma}$ refers to the total low-energy cross section (i.e.\ without the form factor suppression). Since the Sun contains many different elements, $\rho(r)$ in (\ref{re-expr-capture}) should be viewed as the density of a specific element and then summed over all relevant elements in the Sun, with the form factor removed for hydrogen.

Regarding the cross sections, WIMPs can typically scatter via spin-independent and spin-dependent scatterings. The spin-independent scatterings occur on all elements in the Sun, whereas the spin-dependent only occur for nuclei with spin. The only element with spin with non-negligible abundance in the Sun is hydrogen. In Eq.~(\ref{re-expr-capture}), the cross section $\tilde{\sigma}$ refers to the total cross section on a given element in the Sun (without the form-factor suppression). Usually one relates this to the cross section on a single proton, which we will denote $\sigma_{\rm SD}$ and $\sigma_{SI}$ for the spin-dependent and spin-independent cases respectively.
For the spin-dependent case $\tilde{\sigma}=\sigma_{\rm SD}$ (as only scatterings on hydrogen are relevant), whereas for the spin-independent case, the cross sections are related via
\begin{equation}
\tilde{\sigma}=\sigma_\mathrm{SI}A^2
\left(\frac{Mm}{M+m}\right)^2\left(\frac{Mm_\mathrm p}{M+m_\mathrm p}\right)^{-2} \label{si-cross}
\end{equation}
where $m_\mathrm p$ is the mass of the proton and $A$ is the atomic number of the target nucleus.

\begin{figure}
\centerline{\epsfig{file=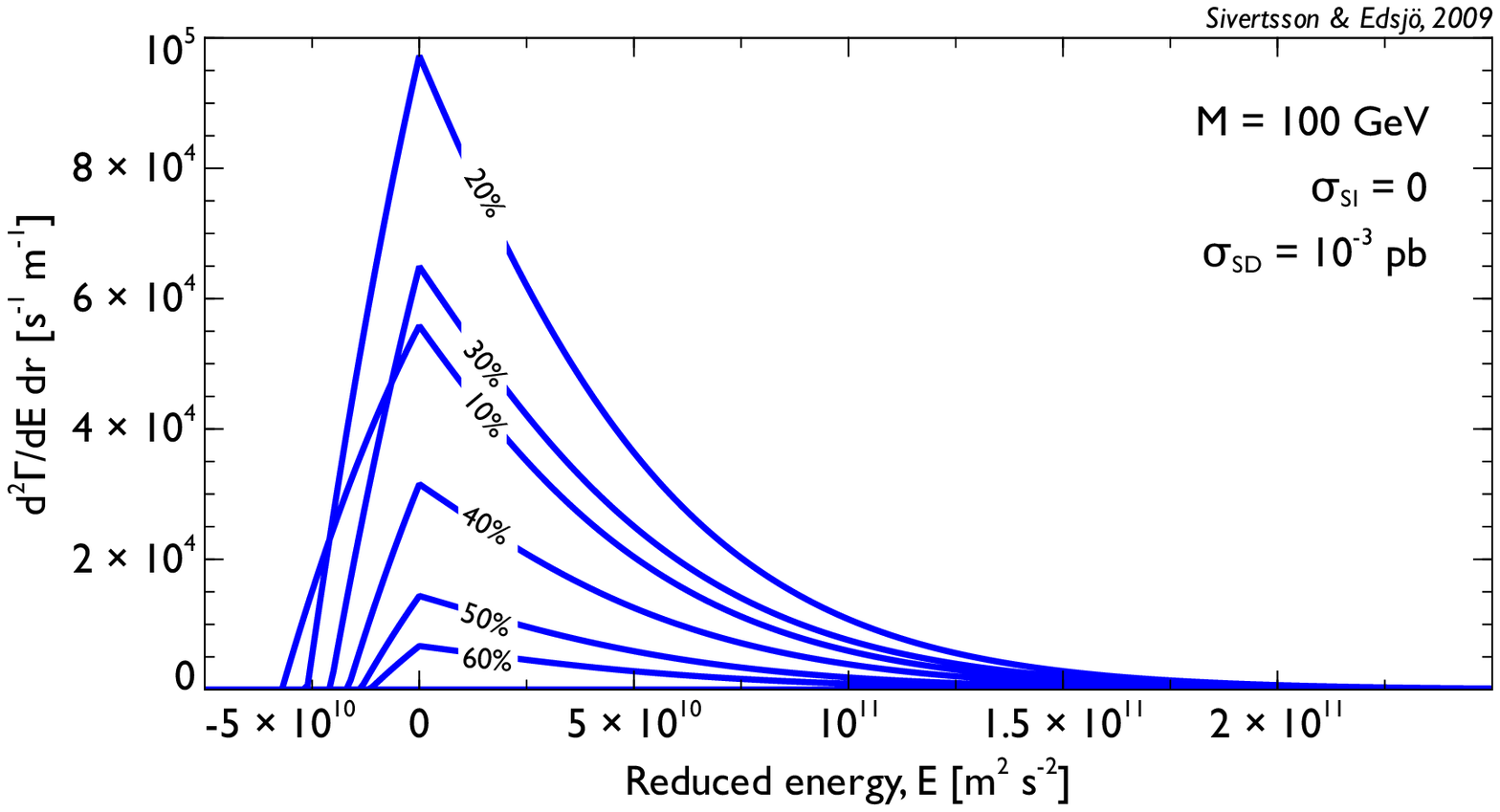,width=0.7\textwidth}}
%\input{plotsdscatter}
%\label{fig:sdscatter}
%\end{figure}
%\begin{figure}
\centerline{\epsfig{file=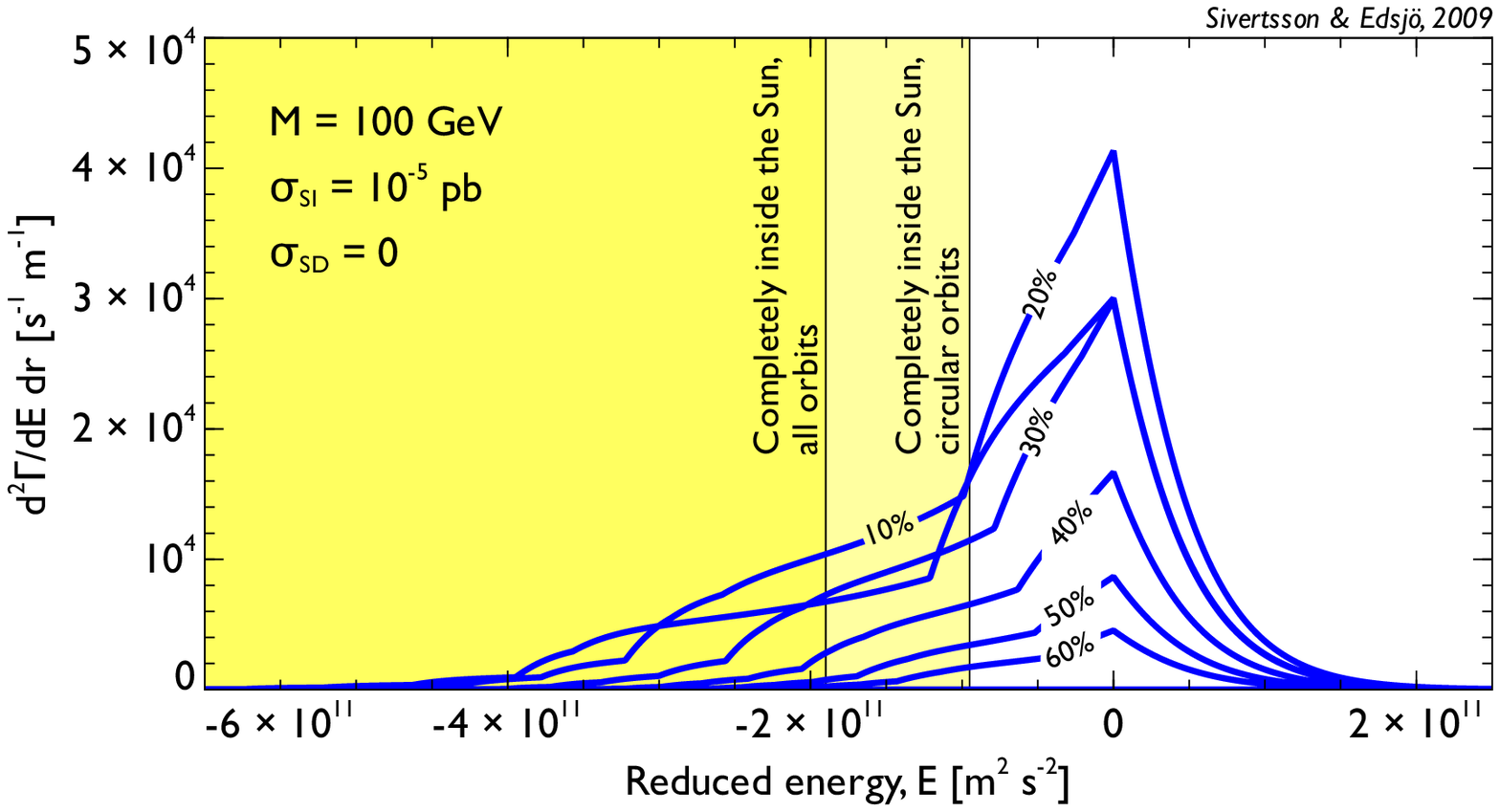,width=0.7\textwidth}}
\caption{Plot of Eq.~(\ref{re-expr-capture}) for the case where spin-dependent scatter, $\sigma_\mathrm {SD}=10^{-3}$ pb, and spin-independent scatter, $\sigma_\mathrm {SI}=10^{-5}$ pb, respectively being dominant.
The graphs show the distribution of scattered WIMPs after their first scatter in the Sun, expressed as the number of scattered WIMPs per unit time, unit reduced energy after the scatter and unit radius in the Sun. The different lines correspond to different radii in the Sun where the WIMPs scatter, showing the shells with radius 10\%, 20\%, \ldots, 60\% of the Sun's radius, the graphs hence illustrates surfaces. The smaller the radius of the illustrated shell, the further to the left the graph has non-zero values, i.e.\ the smaller the possible reduced energy is after the scatter. Also indicated are the reduced energies corresponding to orbits completely trapped in the interior of the Sun.
%In both graphs the highest reaching curve corresponds to the shell at 20\% of the solar radius. 
The WIMP mass is here set to $100$ GeV.}
\label{fig:scatter}
\end{figure}

Eq.~(\ref{re-expr-capture}) describes the distribution of WIMPs after their first scatter in the Sun, illustrated in Fig.~\ref{fig:scatter}. Hydrogen is the only significant element in the Sun which couples axially to the WIMP so if spin-dependent scatter, $\sigma_\mathrm{SD}$, dominates the WIMPs will scatter effectively only off hydrogen. If spin-independent interactions dominate, scatter off hydrogen is insignificant and the WIMPs will essentially only scatter off helium and heavier elements in the Sun. Heavier elements can play an important role even though their abundance is small since the scatter cross section increases for large nuclei and because of the increase of phase space accessible in the scatter. As the WIMPs scatter off nuclei of different masses in the two scenarios the dynamics is quite different. Again, note that both $\sigma_\mathrm{SD}$ and $\sigma_\mathrm{SI}$ refer to the scattering cross sections off protons (as these are the quantities usually used to quantify the size of the scattering cross sections). We will use optimistic values of $\sigma_\mathrm{SI}=10^{-5}$ pb and $\sigma_\mathrm{SD}=10^{-3}$ pb, but as we will see in section \ref{sec:sigmadep}, our results do not depend on the values chosen for the cross sections (within reasonable limits).

In the upper graph in Fig.~\ref{fig:scatter} the WIMPs scatter only off hydrogen which makes the energy loss in the scatter quite small, making the Sun only capable of capturing the low energy tail of the Milky Way WIMP population. This is seen in the graph in that the majority of the WIMPs scatter to an energy greater than zero. As we will discuss later, the WIMPs have to scatter many times before being fully trapped inside the Sun, allowing the bound WIMPs to give larger contributions to the WIMP density outside the Sun.

In the lower graph in Fig.~\ref{fig:scatter} the spin-independent scatter dominate making the WIMPs scatter mostly off heavier nuclei and hence, on average, lose a lot more energy in the scatter. This makes the WIMPs much more likely to be captured in the Sun's potential well but as seen in the graph some will immediately disappear inside the Sun since a WIMP needs a reduced energy of at least $\mathcal E=-\frac{GM_\odot}{R_\odot}=-1.9\times 10^{11}$~m$^2/$s$^2$ to be energetic enough to reach the region outside of the Sun. For circular orbits, $\mathcal E=-\frac{GM_\odot}{2R_\odot}=-0.95\times 10^{11}$~m$^2/$s$^2$ is the limiting energy for the orbits being fully trapped inside the Sun. These two regions are shown in the lower panel in Fig.~\ref{fig:scatter}. As the orbit of a bound WIMP must contain the point of last scatter (up to the spherical symmetry), the WIMP orbits after first scatter are necessarily radial rather than circular. With this in mind we see that it is not very likely that the WIMP loses enough energy to be completely trapped already in the first scatter. For heavier masses than the $M=100$ GeV shown in Fig.~\ref{fig:scatter}, it is even harder to fully capture the WIMP with only one scatter. For spin-dependent scatters (upper panel), not enough energy can be lost for the WIMPs to be completely captured in the Sun with only one scatter.

The pointiness of the graphs in Fig.~\ref{fig:scatter} comes from the discontinuity of the integrand which is related to the footnote associated to Eq.~(\ref{energy_int}). The function $\frac{f_{\mathcal E}(\mathcal E)}{\mathcal E}\theta (\mathcal E)$ jumps from zero to its maximum value at $\mathcal E=0$.

WIMPs which scatter close to the centre of the Sun have higher velocities in the scatter and hence normally lose a larger fraction of their total energy in the scatter. This is the reason for inner shell curves in Fig.~\ref{fig:scatter} being non-zero for lower values of $\tilde\mathcal E$. 

Integrating Eq.~(\ref{re-expr-capture}) over $r$ and $\tilde\mathcal E$, for $\tilde\mathcal E <0$, gives the total capture rate of WIMPs by the Sun. The total capture rate is of interest in computing the neutrino flux from annihilating WIMPs in the centre of the Sun. This can also be calculated using the DarkSUSY program \cite{darksusy} and is in excellent agreement with our results. 

%%%%%%%%

\subsubsection{The scatter angle after the WIMPs' first scatter}
To determine the orbit of the captured WIMP one, besides the energy distribution and the place of scatter, also needs the scatter angle, $\alpha$, defined as the angle between the path of the scattered WIMP and the radial direction from the centre of the Sun.

In this work spherical symmetry is assumed; the Sun is spherically symmetric and isotropy in the incoming WIMPs is assumed. The small scatter probability for a WIMP passing through the Sun makes it equally likely for the WIMP to scatter on its way out as on its way in to the Sun. Because of this the scattered WIMPs inherit the initial WIMP isotropy and all the directions of the just scattered WIMPs are equally probable. If one is to be very careful, the Sun's velocity and gravitational influence could give rise to some small anisotropy.  As the orbits are not purely elliptical inside the Sun, they will precess, which will somewhat smoothen out anisotropies. Gravitational interactions with the planets will also have a similar effect on orbits reaching further out from the Sun. 

All scatter directions being equally probable after the first scatter gives the probability distribution for $\alpha$
\begin{equation}
\frac{2\pi\sin{(\alpha)} \ud\alpha}{4\pi}=\frac{1}{2}\sin(\alpha) \ud\alpha, \mbox{ \ where \ } 0\leq\alpha\leq\pi. \label{alpha-distr}
\end{equation}

%%%%%%%%%%
\subsection{The orbits and lifetimes of the captured WIMPs}
The reduced energy, $\mathcal E$, and reduced angular momentum, $\mathcal J$ (defined as the angular momentum divided by the WIMP mass $M$), of a captured WIMP are given by
\begin{eqnarray}
\mathcal E &=& \frac{v^2}{2}+\mathcal E_\mathrm p(r)\label{orbit_energy}, \\
\mathcal J= rv|\sin\alpha| &=& r\sqrt{2(\mathcal E-\mathcal E_\mathrm p(r))}|\sin\alpha|\label{orbit_angular},
\end{eqnarray}
where $r$ is the radius where the scatter took place and $v$ the velocity the WIMP scattered to. 
The reduced potential energy at the scatter, $\mathcal E_\mathrm p$, is given by a numerical function for the gravitational potential inside the Sun.

The energy and angular momentum together fully specify the orbit, up to the spherical symmetry. It will be useful for us to find the minimal and maximal distances of the orbit to the centre of the Sun, $r_{\min}$ and $r_{\max}$. These two points are the only points in the orbit fulfilling $\mathcal J=r\sqrt{2(\mathcal E-\mathcal E_\mathrm p)}$. We are here only interested in WIMPs which spend time both inside and outside the Sun, i.e.\ WIMPs on orbits fulfilling $r_{\min}<R_\odot <r_{\max}$. As the WIMPs were captured by scatters in the Sun their future orbits will always be partly inside the Sun unless disturbed in some way, e.g.\ by the planets; this is not taken into account in these calculations. Outside the Sun the gravitational potential is simply $\mathcal E_\mathrm p=-GM_\odot /r$, allowing us to simply solve for $r_{\max}$
\begin{equation}
r_{\max}=-\frac{GM_\odot }{2\mathcal E} +\sqrt{\left({\frac{GM_\odot }{2\mathcal E}}\right) ^2 +\frac{\mathcal J^2}{2\mathcal E}}. \label{r_max} 
\end{equation}
Inside the Sun the gravitational potential is not so simple and hence $r_{\min}$ can not be found in such a simple way as $r_{\max}$. The orbit everywhere fulfils $\mathcal J^2\leq (rv)^2=2r^2(\mathcal E-\mathcal E_\mathrm p(r))$ with equality only for $r$ equal to $r_{\min}$ or $r_{\max}$. To find $r_{\rm min}$, we can use that $\mathcal J^2>2r^2(\mathcal E-\mathcal E_\mathrm p(r))$ for $r<r_{\rm min}$ and perform a numerical binary search for $r_{\rm min}$.

%For the evaluation of the rather unnatural points inside the orbit, $r<r_{\min}$, one can show that $\mathcal J^2>2r^2(\mathcal E-\mathcal E_\mathrm p(r))$. It is then easy to determine $r_{\min}$ from a numerical binary search.

\subsubsection{The density contribution to the WIMP halo before the WIMP scatters again.} 
To find the density of WIMPs around the Sun one needs to know what fraction of its time an orbiting WIMP spends at different distances from the Sun. The density contribution from a WIMP at a certain radial interval is proportional to $2\delta t/T$, where $\delta t$ is the time it takes the WIMP to traverse the shell and $T$ is the period of the orbit. The factor of two is added as in completing an orbit the WIMP passes all the radii in the orbit twice. 

We will later look closer at the radial velocities, and from Eq.~(\ref{velocity-over-radial_velocity}) one finds, for the Keplerian potential outside the Sun, that the time it takes to traverse a shell of thickness $|\ud r|$ is
\begin{equation}
\ud t=\frac{r}{\sqrt{2(GM_\odot+\mathcal Er)r-\mathcal J^2}}|\ud r|\label{expr:ellipse-time/radius}.
\end{equation}
This expression diverges for $r=r_{\min}$ and $r=r_{\max}$ but the integral converges everywhere. Since the velocity is lower in the outer regions of the orbit the WIMP also spends most of its time there.

To reduce noise in extracting the density from the Monte Carlo the density contribution from a bound WIMP is given as the integral of the density contribution over the (small) radii interval which is the bin width in the density plots. This is easily done as the integral of Eq.~(\ref{expr:ellipse-time/radius}) can be found in closed form; using {\textit{Mathematica}} a primitive function of the right hand side of Eq.~(\ref{expr:ellipse-time/radius}) is given by
\begin{eqnarray}
-2\frac{GM_\odot}{(-2\mathcal E)^{3/2}}\arctan\left(\frac{2\mathcal Er+GM_\odot}{\sqrt{-2\mathcal E}\sqrt{2r(GM_\odot+\mathcal Er)-\mathcal J^2}}\right)+\frac{1}{\mathcal E}\sqrt{2r(GM_\odot+\mathcal Er)-\mathcal J^2}
\label{integral}.
\end{eqnarray}
As discussed above a factor of 2, compared to Eq.~(\ref{expr:ellipse-time/radius}), has also been added to the primitive function since an orbiting WIMP passes a radial interval twice or not at all.

For the orbits investigated here the WIMP typically spends most of its time outside the Sun. The total time, $T$, it takes for the WIMP to complete a full revolution in an elliptical orbit is
\begin{equation}
T=\frac{\pi}{\sqrt{2}}\frac{GM_\odot}{(-\mathcal E)^{3/2}}.\label{orbit_time}
\end{equation}

The density contribution also depends on how many passages the WIMP survives before it scatters again. For each solar passage there is some probability for the WIMP to scatter in the Sun, reducing the average density contribution from a given bound orbit for each solar passage. The probability for the WIMP to scatter twice in one solar passage is immensely small and can be neglected. 

The probability for the bound WIMP to survive $n$ solar passages without scattering in the Sun is $q^n=(1-P)^n$ with $P$ being the probability for the WIMP to scatter in one passage. The total density contribution by a WIMP in a given orbit, integrated over time, is, on average
\begin{equation}
\rho_0(1+q+q^2+\cdots+q^n) = \rho_0\frac{1-q^{n+1}}{1-q} = \rho_0\frac{1-(1-P)^{n+1}}{P} \label{lifetime},
\end{equation}
where $\rho_0$ is the density contribution from the WIMPs first revolution in the given orbit. The finite age, of approximately 4.5 billion years, of the Sun gives a cut in the sum since no orbits can be older than this. The maximum number of revolutions a WIMP can have fulfilled in a given orbit is $n = (4.5\times 10^9\mbox{\ yr})/T$, with $T$ being the orbit time, as in expression~(\ref{orbit_time}). 

In practice, the finite age of the Sun is important only for very low scatter cross sections. Then also the orbits which these long lived WIMPs scatter to should be removed. This was taken into account in our runs with very low scatter cross section.

\subsubsection{The scatter probability for a bound WIMP passing through the Sun}
The probability for collision when a WIMP passes a shell in the Sun is given by
\begin{eqnarray}
\sigma\frac{\rho (r)}{m}v\times 2\frac{dr}{|v_\mathrm r|}\label{expr-hitprob},
\end{eqnarray}
where $\rho$ and $m$ refer to the density and mass, respectively, of the target nucleus. The above expression is then summed over the relevant elements. Also, the scatter cross section can depend on the energy loss in the scatter, which depends on the WIMP velocity and hence the radius of the scatter.
The above expression comes from the number of scatters per unit time $(\sigma nv)$ times the time the WIMP spends in a given shell taken twice, since the WIMP intersects a shell twice or not at all in one solar passage.

From $v$ and $\mathcal J$ for a particle one easily determines the radial velocity component
\begin{eqnarray}
\frac{v}{|v_\mathrm r|} &=& \frac{rv}{\sqrt{r^2v^2-\mathcal J^2}}. \label{velocity-over-radial_velocity}
\end{eqnarray}
Hence, the probability for the bound WIMP to scatter in the Sun in one solar passage is
\begin{eqnarray}
P &=& \frac{2}{m}\int_{r_{\min}}^{R_\odot}\sigma \rho(r)\frac{rv}{\sqrt{r^2v^2-\mathcal J^2}}\,\ud r, \label{scatterprob}
\end{eqnarray} 
where $v^2=2\mathcal E-2\mathcal E_\mathrm p(r)$ and $r_{\min}$ is easily determined numerically, as previously discussed. If the WIMP scatters off hydrogen, the cross section does not suffer a form-factor suppression and it is straightforward to evaluate Eq.~(\ref{scatterprob}). 

Accounting for the form factor suppression of the cross section needs some further treatment. For a bound WIMP passing the Sun once in an orbit with reduced energy, $\mathcal E$, and reduced angular momentum, $\mathcal J$, the probability to scatter off a given element $i$ in a (possible) shell in the Sun, given by the radius $r$, to a certain, possible reduced energy $\tilde{\mathcal E}$ is given by
\begin{equation}
\frac{\ud^2 P_i}{\ud r \ud \tilde{\mathcal E}} = 
\frac{1}{\mathcal E_\mathrm k}\frac{(M+m_i)^2}{4Mm_i} e^{-\Delta \mathcal E/\mathcal E_0^i} \ \tilde{\sigma_i} \frac{\rho_i(r)}{m_i} \frac{2}{v_\mathrm r}, \label{shell-hitprob-form}
\end{equation}
where the rightmost part is simply Eq.~(\ref{expr-hitprob}). The first part of Eq.~(\ref{shell-hitprob-form}) is the factor given in Eq.~(\ref{expr-probability}) times the form factor suppression.

The probability for the orbiting WIMP to scatter to any possible energy in a given shell during one solar passage is given by the integral of Eq.~(\ref{shell-hitprob-form}) over the possible values of $\tilde{\mathcal E}$ and then summed over the different elements in the Sun
\begin{equation}
\frac{\ud P}{\ud r} = 
\frac{v}{v_\mathrm r}\frac{1}{\mathcal E_\mathrm k}\left(2(\sigma_\mathrm{SD}+\sigma_{\rm SI})\frac{\rho_1(r)}{m_1}+
\sum_{i=2}^{16}\frac{(M+m_i)^2}{2Mm_i}\tilde{\sigma}_\mathrm{SI}^i\frac{\rho_i(r)}{m_i}\mathcal E_0
\left[1-\exp\left(-\frac{4Mm_i}{(M+m_i)^2}\frac{\mathcal E_\mathrm k}{\mathcal E_0}\right)\right]\right), \label{shellhitprob}
\end{equation}
where $v/v_\mathrm r$ is given by Eq.~(\ref{velocity-over-radial_velocity}). The mass and density of element $i$ are given by $m_i$ and $\rho_i(r)$ respectively, with $i=1$ being hydrogen. The leftmost term refers to scattering off hydrogen, which is the only element here with spin-dependent scatter cross section to the WIMP (the spin-independent scattering off hydrogen is added to this term as well for clarity). 
$\tilde{\sigma}_\mathrm{SI}^i$ here refers to the total low-energy spin-independent cross section on element $i$, and is given by $\tilde{\sigma}$ in Eq.~(\ref{si-cross}).

The probability for the captured, bound WIMP to scatter during one solar passage in its orbit is then given by the integral of the above expression over the radii in the Sun that the WIMP passes through in its orbit.

\subsubsection{The density contribution of the WIMPs which have only scattered once}

\begin{figure}
\centerline{\epsfig{file=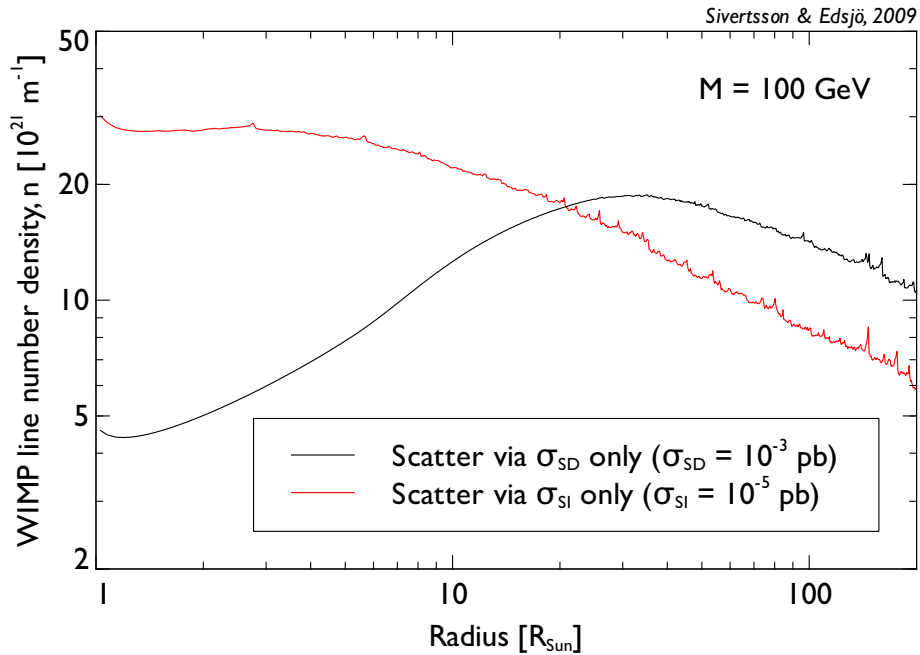,width=0.7\textwidth}}
\caption{The line density $n$ around the Sun of WIMPs that have scattered exactly once in the Sun. The black curve is for spin-dependent scatters dominating, using $\sigma_\mathrm{SD}=10^{-3}$~pb, $\sigma_\mathrm{SI}=0$ and is generated by a calculation of 1~million scattered WIMPs. The red curve is for spin-independent scatters dominating, using $\sigma_\mathrm{SI}=10^{-5}$~pb, $\sigma_\mathrm{SD}=0$ and is the result of 100\,000 simulated scattered WIMPs. For both curves the WIMP mass is $M=100$~GeV. The Earth is approximately 215 solar radii away from the Sun.}
\label{fig:montefirst}
\end{figure}

The analysis is now sufficient to calculate the density around the Sun of WIMPs which have scattered exactly once in the Sun. This is done through making a Monte Carlo simulation, simulating a large number of captured WIMPs according to the energy distribution after their first scatter, given by Eq.~(\ref{re-expr-capture}). This is described in the first part of section \ref{monte-carlo}, which describes the structure of the final full Monte Carlo. The resulting density contribution of the WIMPs which have scattered only once is shown in Fig.~\ref{fig:montefirst}. This graph shows the line number density, $n$, of WIMPs around the Sun. The line number density is the number of WIMPs in a spherical shell of a given radius $r$ per radial interval and it is related to the three-dimensional spatial density $N$ (number of WIMPs per volume element) as
\begin{equation}
N(r) = \frac{n(r)}{4\pi r^2} \label{eq:linenumberdensity}
\end{equation}
The line number density in Fig.~\ref{fig:montefirst} is the contribution to the line number density from the WIMPs that have scattered once, but not yet scattered a second time, and become gravitationally bound to the Sun. As they scatter more times, they will move in closer to the Sun and the contribution will increase at lower radii. The reason the spin-dependent curve peaks at higher radii is kinematical as the WIMPs in this case have only scattered on hydrogen. The spin-independent curve includes scatterings on heavier elements, which causes the WIMPs to lose more energy in the scattering.

The unevenness in Fig.~\ref{fig:montefirst} is due to the stochastic nature of Monte Carlo simulations. We have simulated a finite number of WIMP orbits to calculate the contribution to the line number density and as some simulated orbits can be quite stable (meaning that they have a long life time and hence a large contribution to the line number density), we can get large contributions from a few orbits. The peaks then occur at the turn-around radius $r_{\rm max}$, where the WIMPs spend most of their time and these are the peaks we see in the figure. The more WIMP orbits we simulate, the more even the curves get, but given the fact that we will be most interested in the density close to the Sun (where the density is highest, and the curves are smoother), we do not have to worry about peaks at large radii. Also, the spin-dependent curve is smooth at small radii as in this case the WIMPs close to the Sun are never close to the turn-around radius $r_{\max}$ after the first scattering.

Taking into account the fact that the orbits are not perfect ellipses inside the Sun is found to increase the calculated line number density by roughly $20\%$. 
This is because the true orbits do not come as far into the dense central regions in the Sun as purely elliptical orbits would, which increases the life time of the true orbits since the scatter probability decreases. 

%%%%%%%%%%
\subsection{Subsequent scatters in the Sun} \label{section:further_scatters}
In general a WIMP scatters many times before its orbit becomes completely trapped inside the Sun. These intermediate orbits contribute to the WIMP density, especially in the region close to the Sun.

The radius of the point of scatter for an already bound WIMP and the type of nucleus involved can in the Monte Carlo be extracted from Eq.~(\ref{shellhitprob}), since this encodes the distribution of the scatter parameters for a given orbit. Knowing the place of scatter and the elements involved, the distribution of the energy loss in the scatter is found by expression~(\ref{shell-hitprob-form}).

To find the WIMP's angular momentum after the scatter, $\tilde J$, requires some further treatment. Once the properties of the new orbit are determined the contribution to the halo density can be calculated in the same way as for the WIMPs which have only scattered once.

Using conservation of energy and momentum in the scatter and the general relation $p^2=\mathbf p\cdot \mathbf p=2ME_\mathrm k$ one can derive the relation
\begin{equation}
\mathbf p\cdot \tilde\mathbf p=(M-m)E_\mathrm k+(M+m)\tilde E_\mathrm k, \label{p_dot_p_tilde}
\end{equation}
where $\mathbf p$ and $\tilde\mathbf p$ are the momentum before and after the scatter, respectively.
Dividing $\tilde\mathbf p$ and $\mathbf r$ in components parallel ($\tilde p_\parallel$ and $r_\parallel$) and orhogonal ($\tilde\mathbf p_\perp$ and $\mathbf r_\perp$) to $\mathbf p$, using (\ref{p_dot_p_tilde}) and $\mathbf p\cdot \tilde\mathbf p=pp_\parallel$ gives
\begin{equation}
\tilde p_\parallel=\frac{(M-m)E_\mathrm k+(M+m)\tilde E_\mathrm k}{p} \mbox{ \ \  \ and \ \ \ } \tilde p^2_\perp=\tilde p^2-\tilde p^2_\parallel.
\end{equation}
From the angular momentum before the scatter, $J=|\mathbf J|=r_\perp p$, we can derive $r_\perp$ and $r_\parallel$ since both $p$ and $J$ are known.

The modulus of the angular momentum after the scatter can be evaluated as
\begin{equation}
|\tilde\mathbf J|^2=
\left|\left(\begin{array}{c} 
r_\perp\\0\\r_\parallel\\
\end{array}\right)
\cdot \left(\begin{array}{c} \tilde p_\perp\cos\phi\\
\tilde p_\perp\sin\phi\\ \tilde p_\parallel\\ \end{array}\right)\right|^2=
r^2\tilde p_\perp^2 \sin^2\phi+ (r_\parallel \tilde p_\perp\cos\phi - r_\perp\tilde p_\parallel)^2.
\end{equation}
The azimuthal angle $\phi$ is arbitrary in any given scatter and is hence generated as a random number in the Monte Carlo simulation. 

%%%%%%%%%%
\subsection{Monte Carlo} \label{monte-carlo}

The actual calculation of the density of the Sun's WIMP halo is implemented through constructing a Monte Carlo simulating the capture process of a large number of WIMPs. The Monte Carlo is constructed as follows.
\begin{enumerate}
\item\label{list:inside} The Monte Carlo starts from the distribution of WIMPs which have scattered once to bound orbits. This distribution of WIMP interaction points, energy after the scatter and scatter element (from the solar model) is given by Eq.~(\ref{re-expr-capture}); this distribution is also shown in Fig.~\ref{fig:scatter}. Having picked a WIMP according to this distribution, the scatter angle is picked from Eq.~(\ref{alpha-distr}), giving the final piece of information required to calculate the WIMPs angular momentum. 
\item\label{list:outside} The density contribution by the WIMP in the orbit, before it scatters again, is then calculated. The line density contribution in a radial interval for one revolution in the orbit is given by Eq.~(\ref{integral}). This is then multiplied with the average number of solar passages that a WIMP in such an orbit survives, which is given by Eq.~(\ref{lifetime}) and Eq.~(\ref{scatterprob}).
\item The shell in the Sun in which the WIMP scatters next time is picked, using the expression for the scatter probability in the passage of a shell, given by Eq.~(\ref{shellhitprob}). The type of element the WIMP scatters off in the shell is then chosen according to the elemental abundance in the Sun at that radius.
\item The WIMPs energy loss in the scatter and its new angular momentum is determined, as described in section \ref{section:further_scatters}.
\item If the WIMP's new orbit is still energetic enough to be partly outside the Sun ($r_{\max}>R_\odot$ in Eq.~(\ref{r_max})), the simulation continues with step \ref{list:outside}, continuing to add to the WIMP density around the Sun.
\item If the WIMP has lost enough energy to be completely trapped inside the Sun it is no longer relevant here. The simulation of a new scattered WIMP starts from step \ref{list:inside}.
\item When the desired number of captured WIMPs are simulated the result is weighted with the capture rate, which is found by integrating Eq.~(\ref{re-expr-capture}) over $r$ and $\tilde\mathcal E$ for $\tilde\mathcal E<0$, to get the real density.
\end{enumerate}

\section{The WIMP number density in the halo}
\begin{figure}
\centerline{\epsfig{file=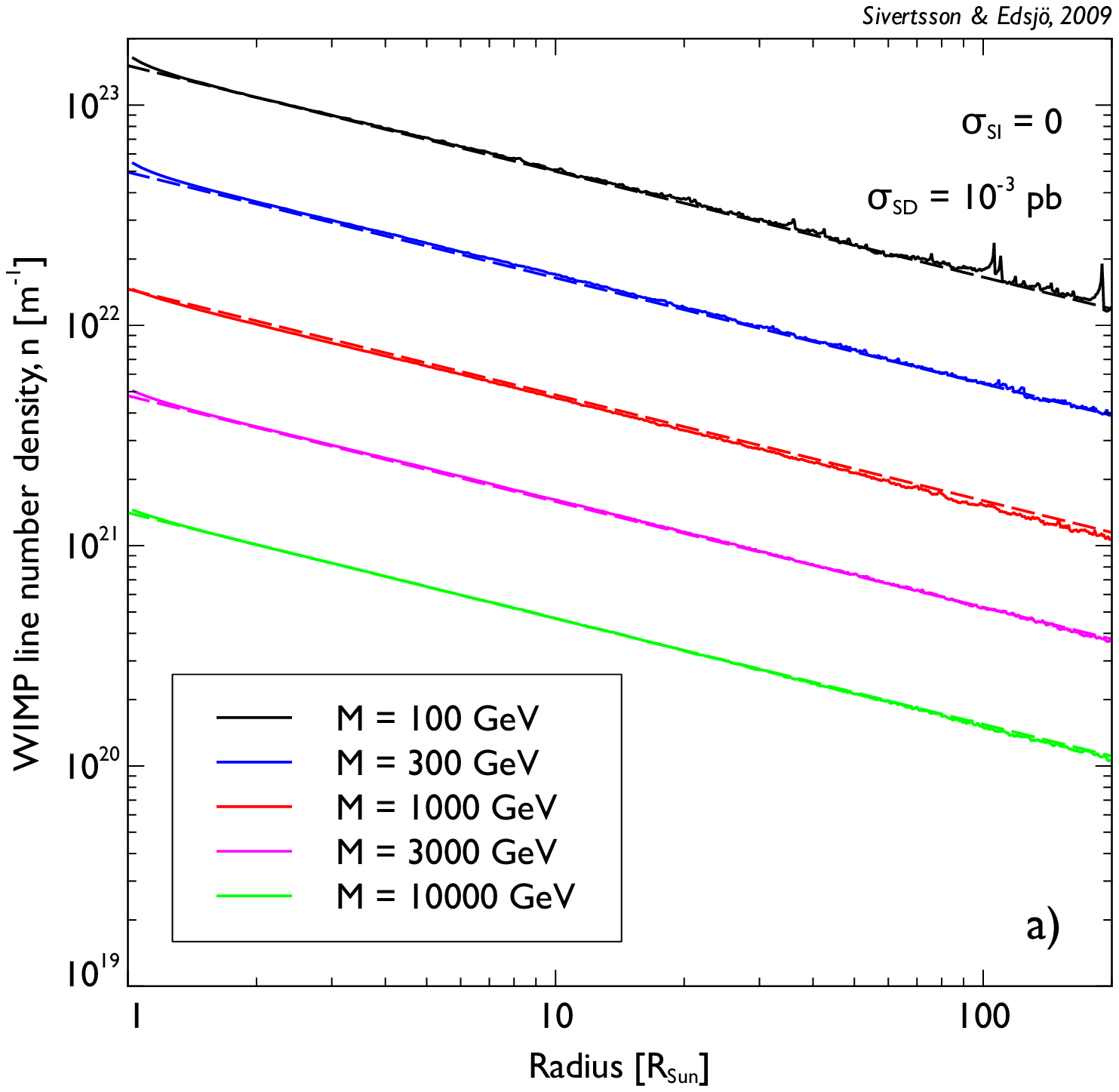,width=0.49\textwidth}\hfill \epsfig{file=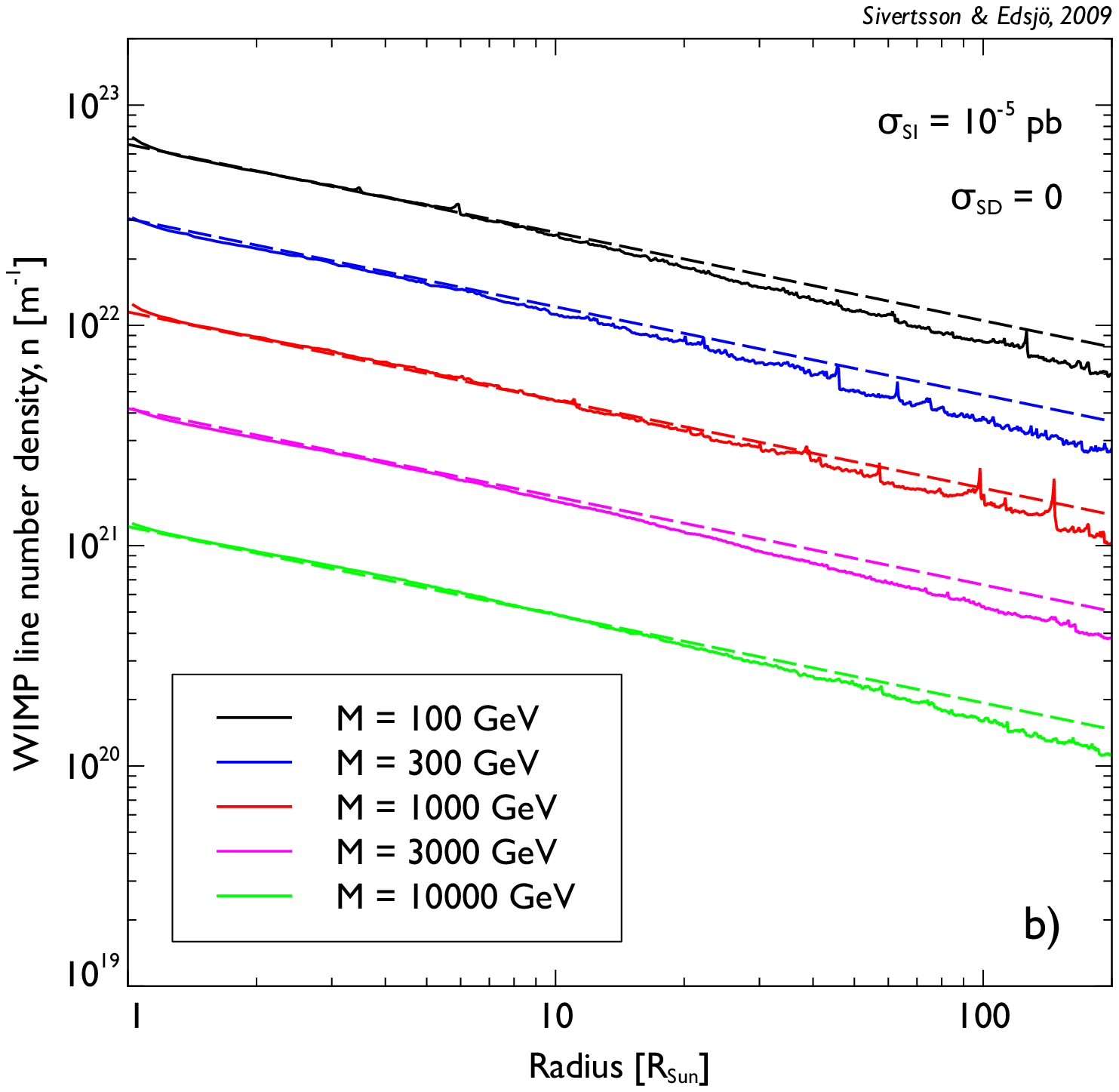,width=0.49\textwidth}}
\caption{a) The line number density of WIMPs around the Sun with spin-dependent cross section being dominant, i.e.\ WIMPs do only scatter off hydrogen. In this figure $\sigma_\mathrm{SD}=10^{-3}$ pb and $\sigma_\mathrm{SI}=0$ pb. The dashed curves are the fits described by expression~(\ref{eq:nsd}). b) The line number density of WIMPs around the Sun with spin-independent cross section being dominant, i.e.\ WIMPs scatter off all elements in the Sun. In this figure $\sigma_\mathrm{SD}=0$ pb and $\sigma_\mathrm{SI}=10^{-5}$ pb. The dashed curves are the fits described by expression~(\ref{eq:nsi}).}
\label{fig:finalsdsi}
\end{figure}

The constructed Monte Carlo has, as described in the above section, simulated the capture process, giving the density of WIMPs around the Sun. The simulations have been made for different WIMP masses and for both spin-dependent and spin-independent cross section being dominant. In total we have simulated about $10^7$ WIMPs from first scatter to complete solar entrapment. The results of these simulations are shown in Fig.~\ref{fig:finalsdsi}. The line number density can be well approximated by
\begin{eqnarray}
n^{SD} & = & 10^{25.21 - 1.015x} \left(\frac{r}{R_\odot}\right)^{-0.48} \mbox{ m$^{-1}$},\label{eq:nsd}\\
n^{SI} & = & 10^{23.71 - 0.2332x -0.1056x^2} \left(\frac{r}{R_\odot}\right)^{-0.40} \mbox{ m$^{-1}$}, \label{eq:nsi}\\
\mbox{with } x & = & \log_{10} \left( \frac{M}{\mbox{1 GeV}} \right).
\end{eqnarray}
The three-dimensional spatial WIMP number density, $N$, is then given from the line number density by Eq.~(\ref{eq:linenumberdensity}). The fits given above are optimized for small radii, as these are the most important for calculating the flux from annihilation in the solar WIMP halo. Note that the densities we find here are only marginally larger than the background density of WIMPs in the Galactic halo (about a factor of ten larger at the solar surface), so the density enhancements are not very dramatic. 

%%% Figures

%%%%%%%%%%
\subsection{The halo number density dependence on the cross section and WIMP mass}
\label{sec:sigmadep}

The magnitude of the cross section, except for very low cross sections, is insignificant for the WIMP density around the Sun. This is due to the capture rate of Galactic WIMPs in the Sun being proportional to the cross section, while the lifetime of a WIMP orbit is proportional to the inverse of the cross section. This cancels the cross section dependence if equilibrium has been reached, which happens if the Sun's capture rate of WIMPs from the Galactic halo equal the rate at which captured WIMPs become fully trapped inside the Sun. For very low scatter cross sections this process becomes so slow for a substantial fraction of the WIMPs that the age of the Sun is not long enough for equilibrium to have been reached, reducing the WIMP density around the Sun for very low scatter cross sections. The internal relation between the spin-dependent and spin-independent cross section is however important, since this determines whether or not the WIMPs scatter mostly off hydrogen or mostly off heavier elements. 

The WIMP mass density ($\rho=n M$) around the Sun is seen to not be all that sensitive to the WIMP mass; Fig.~\ref{fig:finalsdsi} shows the WIMP (line) number density which depends on WIMP mass, as does the WIMP number density in the Milky Way WIMP halo. For heavy WIMPs the mass difference between the WIMP and the target nucleus is larger, making it more unlikely for a Galactic halo WIMP to lose enough energy in a scatter to become gravitationally bound to the Sun. On the other hand a heavy WIMP also requires more scatters to lose enough energy to be completely trapped inside the Sun, allowing each captured WIMP to contribute to the dark matter density around the Sun for a longer time. These two effects somewhat cancel, making the WIMP mass density largely insensitive to WIMP mass. Note that as heavier WIMPs lose their energy more slowly, the WIMP density in this case becomes slightly more sensitive to very low scatter cross sections. 

%%%%%%%%%%
\section{Gamma-ray fluxes at Earth}

\subsection{Signal fluxes from WIMP annihilations in the solar halo}

The number of WIMP annihilations, $\Gamma_\mathrm A$, for a given volume is in general given by
\begin{equation}
\Gamma_\mathrm A=\frac{1}{2}\langle\sigma v\rangle \int n^2 \,\ud V. \label{general_gamma}
\end{equation}
For simplicity, let us assume a standard annihilation cross section of $\langle \sigma v\rangle=3 \times 10^{-26} \mbox{  cm}^3\mbox{s}^{-1}$, that gives about the correct relic density of WIMPs.

The flux of gamma radiation at Earth from a small source is
\begin{equation}
\phi_\gamma=\Gamma_\mathrm A\frac{1}{4\pi l^2}N_\gamma, \label{general_flux}
\end{equation}
where $l$ is the distance between the gamma-ray source and the telescope at Earth. The average number of gamma-rays created per WIMP annihilation, $N_\gamma$, requires further assumptions about the WIMP candidate investigated. We here assume that the WIMP annihilations create $b\bar{b}$, which give reasonably high fluxes of gamma-rays. With {\sf DarkSUSY 5.0.4} \cite{darksusy}, we have calculated that we then get about $N_\gamma = 14.1, 45.7$ and $94.6$ for $M=100, 1\,000$ and $10\,000$ GeV respectively. These $N_\gamma$ are calculated above 1 GeV in gamma energy. We could also get monochromatic gamma-rays from annihilation to $\gamma\gamma$ and $Z\gamma$ or a harder spectrum from annihilation to e.g.\ $\tau^+ \tau^-$, but (as we will see), given the low fluxes of gamma-rays we predict, it does not make any major difference to our conclusions including these other channels or not.

Combining expressions (\ref{general_gamma}) and (\ref{general_flux}) for our spherically symmetric WIMP distribution, one obtains the observed flux at Earth per solid angle
\begin{equation}
\frac{\ud\Phi_\gamma}{\ud\Omega}=\frac{1}{8\pi}\langle\sigma v\rangle N_\gamma\int n^2(r)\,\ud l, \label{flux_per_steradian}
\end{equation}
which can be evaluated using $r^2=l^2+R^2-2Rl\cos (\theta)$, where $R$ is the distance between the Earth and the Sun, $\theta$ is the angle between the direction of observation and the direction towards the solar centre and $r$ is the distance from the Sun. Since gamma-rays can not pass through the Sun one gets a limit in $l$ at the Sun's surface in the integration, when the line of sight intersects the Sun, giving
\begin{equation}
l\leq R\cos\theta -\sqrt{R_\odot^2-R^2\sin^2\theta} \mbox{ \ \ \ when \ \ \ } \theta \leq \arctan \left(\frac{R_\odot}{R}\right).
\end{equation}

\begin{figure}
\centerline{\epsfig{file=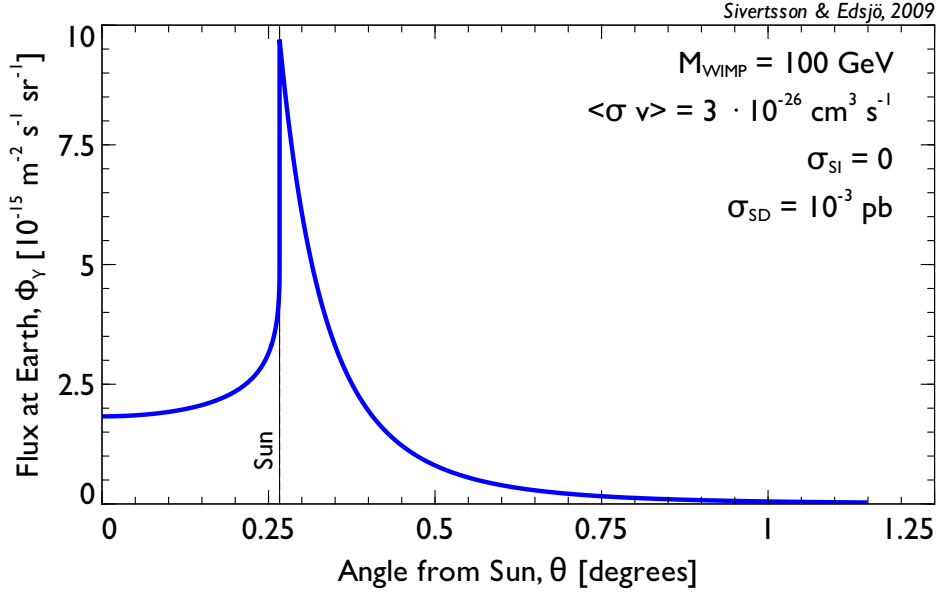,width=0.7\textwidth}}
\caption{The flux per solid angle from the Sun's WIMP halo as seen from Earth for $M=100$ GeV, $\sigma_\mathrm{SD}=10^{-3}$ pb, $\sigma_\mathrm{SI}=0$ and $N_\gamma = 14.1$. Plotted against the angle of sight $\theta$, looking towards the centre of the Sun corresponds to $\theta=0$.}
\label{fig:fluxsr}
\end{figure}

The gamma-ray flux at Earth per solid angle, from Eq.~(\ref{flux_per_steradian}), is shown for a 100 GeV WIMP in Fig.~\ref{fig:fluxsr}. When looking just outside the rim of the Sun the graph peaks since the distance probed by the telescope suddenly increases (as does the astrophysical background). The gamma-ray flux decreases fast with increasing $\theta$ since the WIMP overdensity falls off fast with distance. Note that we have here not included the background density of WIMPs from the Galactic halo, that would actually dominate for the large distances away from the Sun.

The total flux at Earth from these WIMP annihilations integrated out to some maximum angle, $\theta_{\max}$, is 
\begin{equation}
\Phi_\gamma=\frac{1}{4}\langle\sigma v\rangle N_\gamma\int_0^{\theta_{\max}}\int_0^{l_{\max}} n^2(r)\sin\theta\,\ud l \,\ud\theta. \label{total_flux}
\end{equation}

In Table~\ref{table:flux}, we show this flux for some WIMP parameter configurations (integrated over all $\theta$, as the end result is not very sensitive on the actual value of $\theta_{\max}$). One should also note that the WIMP annihilation rate in the solar halo is much lower than the capture rate and hence has no significant impact on the WIMP density in the solar halo.

\begin{table}[ht]
\centering % used for centering table
\begin{tabular}{c|cccccc}
$M$ (GeV) & 100 & 100 & $1\,000$ & $1\,000$ & $10\,000$ & $10\,000$ \\
$\sigma_\mathrm{SI}$ (pb) & 0 & $10^{-5}$ & 0 & $10^{-5}$ & 0 & $10^{-5}$ \\
$\sigma_\mathrm{SD}$ (pb) & $10^{-3}$ & 0 & $10^{-3}$ & 0 & $10^{-3}$ & 0 \\
	\hline
$N_\gamma (>1 \mbox{~GeV})$ & 14.1 & 14.1 & 45.7 & 45.7 & 94.6 & 94.6  \\ \hline
$\Phi_\gamma^{>1\mbox{\scriptsize ~GeV}}$ (m$^{-2}$ s$^{-1}$) & $8.4 \times 10^{-19}$ & $1.8 \times 10^{-19}$ 
& $2.5 \times 10^{-20}$ & $1.7 \times 10^{-20}$
& $4.9 \times 10^{-22}$ & $4.0 \times 10^{-22}$\\ \hline
$F_\mathrm{Milagro}$ (m$^2$ s) & 0 & 0 & $2.2 \times 10^5$ & $2.2 \times 10^5$
& $3.6 \times 10^7$ & $3.6 \times 10^7$ \\
$N_\mathrm{Milagro}$ (events) & 0 & 0 & $5.6 \times 10^{-15}$ & $3.8 \times 10^{-15}$ & $1.8 \times 10^{-14}$ & $1.5 \times 10^{-14}$ \\
$F_\mathrm{HAWC}$ (m$^2$ s) & $2.1\times 10^6$ & $2.1\times 10^6$ & $1.6 \times 10^{9}$ & $1.6 \times 10^{9}$
& $7.9 \times 10^{10}$ & $7.9 \times 10^{10}$ \\
$N_\mathrm{HAWC}$ (events) & $1.7\times 10^{-12}$ & $3.7\times 10^{-13}$ & $3.9 \times 10^{-11}$ & $2.6 \times 10^{-11}$ & $4.0 \times 10^{-11}$ & $3.3 \times 10^{-11}$ \\
$F_\mathrm{ARGO-YBJ}$ (m$^2$ s) & $2.8\times 10^{10}$ & $2.8 \times 10^{10}$ & $3.3 \times 10^{11}$ & $3.3 \times 10^{11}$
& $4.5 \times 10^{12}$ & $4.5 \times 10^{12}$ \\
$N_\mathrm{ARGO-YBJ}$ (events) & $2.4\times 10^{-8}$ & $5.1 \times 10^{-9}$ & $8.2 \times 10^{-9}$ & $5.6 \times 10^{-9}$ & $2.3 \times 10^{-9}$ & $1.9 \times 10^{-9}$ \\
$F_\mathrm{Fermi}$ (m$^2$ s) & $2.2\times 10^{7}$ & $2.2\times 10^{7}$ & $2.3\times 10^{7}$ & $2.3\times 10^{7}$
& $2.3\times 10^{7}$ & $2.3\times 10^{7}$ \\
$N_\mathrm{Fermi}$ (events) & $1.9\times 10^{-11}$ & $4.0 \times 10^{-12}$ & $5.7 \times 10^{-13}$ & $3.9 \times 10^{-13}$ & $1.1 \times 10^{-14}$ & $9.3 \times 10^{-15}$ \\
\end{tabular}
\caption{A set of benchmark WIMP models and their corresponding gamma-ray fluxes above 1 GeV at Earth from WIMP annihilations in the solar halo. Also shown are the integrated effective areas convolved with the gamma-ray spectrum $dN_\gamma/dE$ for WIMP annihilations to $b \bar{b}$, $F$. The total number of events expected in the Milagro search for gamma-rays from the Sun is also shown. For Milagro, the number of events are given for the specific search for gamma-rays from the Sun that has been carried out in \protect\cite{milagro}. For HAWC, ARGO-YBJ and Fermi, 5 years of observation of the Sun has been assumed with the effective areas taken from \protect\cite{hawc-dingus-talk}, \protect\cite{argo} and \protect\cite{fermi-lat-performance}.
The annihilation cross section is assumed to be $\langle \sigma v\rangle = 3\times 10^{-26}$ cm$^{-3}$ s$^{-1}$.}
\label{table:flux}
\end{table}

Given an experiment with an effective area $A$ and live-time $T$, we can write the total number of events (total number of photons) as
\begin{equation}
N_{\rm events} = \Phi_\gamma F
\end{equation}
where
\begin{equation}
F = A T.
\end{equation}
In reality though, the 
effective area $A$ will depend (more or less strongly, depending on experiment) on energy, and a more accurate way of calculating $F$ is to convolve the gamma-ray energy spectrum (where we use $b\bar{b}$ as our template spectrum) with the integrated (over time) effective areas, $A_{\rm eff}^T$,
\begin{equation}
F = \int_0^{M} A_{\rm eff}^T(E) \frac{1}{N_\gamma}\frac{dN_\gamma}{dE} dE \label{eq:Nphotons}
\end{equation}
which is the expression we will use to calculate the number of events for some relevant gamma-ray experiments.

Milagro  has searched for gamma-rays from the Sun and with their estimated integrated effective areas $A_{\rm eff}^T$ for these searches \cite{milagro}, based on 1165 hours of observation of the Sun, we can estimate the total number of events $N_{\rm Milagro}$ these WIMP annihilations should have produced. In Table~{\ref{table:flux}} we give both $F$ and the total number of events for Milagro.
As can be seen, the total number of events are at best in the $10^{-14}$ range, an extremely small number of events, which of course would not have shown up in the Milagro searches at all.

Aside from Milagro, there are other gamma-ray experiments (\textit{Fermi} \cite{fermi}, HAWC \cite{hawc}, and ARGO-YBJ \cite{argo}) that might be sensitive to a gamma-ray excess from the Sun. Air \v Cerenkov telescopes cannot look at the Sun as they can only observe when the sky is dark. \textit{Fermi} has an effective area of slightly below 1 m$^2$ and a field of view of about $\pi$ sr. We have integrated the effective area (integrated over field of view) for \textit{Fermi} \cite{fermi-lat-performance}, which for five years of operation gives a total number of events less than 10$^{-10}$, far below any detection limit. HAWC on the other hand will be much larger than Milagro and \textit{Fermi} and will have an effective area about 100 m$^2$ at 100 GeV and about few times 10$^4$ m$^2$ at higher energies \cite{hawc-dingus-talk}. We have calculated $F$ for HAWC and the corresponding number of events in five years livetime and the number of events are at most in the $10^{-11}$ range, i.e.\ very small. Even though  HAWC is much bigger than {\it Fermi}, it has a rather high threshold of around 30 GeV and quite low effective areas for low energies, which makes the rates rather small for typical WIMP masses.
ARGO-YBJ is designed to have a low energy threshold (sub-GeV) and the effective area increases rapidly with energy. At 1 TeV, the effective area is at most 10$^6$ m$^2$ \cite{argo} (assuming the optimistic 1 multiplicity channel for scaler mode operation). We have calculated $F$ also for ARGO-YBJ given the effective areas for this mode in \cite{argo}. These and the corresponding number of events are also given in Table~\ref{table:flux}. The number of events are at most in the $10^{-8}$ range for five years live-time. All these values for $F$ and the expected number of events are given in Table~\ref{table:flux}. The signals are really diminutive, even with the optimistic assumptions for current and future detectors used here.

\subsection{Backgrounds of gamma-ray fluxes from the Sun}

As the Sun does not shine at such high photon energies and also shields against the diffuse gamma-ray background, one might naively assume that this signal has essentially no background. However, secondary cosmic-ray processes produce gamma-rays from around the Sun. Cosmic-ray protons interacting with the Sun's chromosphere \cite{1991ApJ...382..652S}, produce a gamma-ray flux of about $10^{-4}$ m$^{-2}$ s$^{-1}$ above 1 GeV.

Solar photons can also be up-scattered in energy by cosmic-ray electrons \cite{Orlando:2006zs,Moskalenko:2006ta,2007ApJ...664L.143M}, which produce an even larger flux of gamma-rays than the proton interactions. This gamma-ray flux has also been measured by \textit{EGRET} data \cite{2008A&A...480..847O}, where a flux of about $3 \times 10^{-3}$ m$^{-2}$ s$^{-1}$ in the energy range 100--300 MeV has been seen. Also \textit{Fermi} has observed this gamma flux in the energy range 100~MeV--10~GeV and the preliminary results agree well with the predicted background \cite{Brigida}. One should also note that this background depends on the solar cycle, and as the Sun is more quiet now, we expect a slightly larger astrophysical gamma-ray background at the present \textit{Fermi} result than the older \textit{EGRET} measurement.

Either way, comparing these background fluxes with the expected signal fluxes in Table \ref{table:flux}, we see that the expected signal is indeed, not only very small, but also extremely well hidden in the backgrounds. The proton-induced background is expected to be quite well-defined to the solar rim, whereas the inverse-Compton up-scattered photons are expected to be more extended. In both cases, the Sun itself should appear as a shadow as it blocks the cosmic-rays from the other side. The WIMP signal on the other hand, should arise also from the Sun itself as is seen in Fig.~\ref{fig:fluxsr}. However, even if we expect a different angular distribution and energy spectrum of the signal, it is going to be virtually impossible to disentangle the WIMP signal from the backgrounds, even for an overwhelmingly huge future gamma-ray detector looking towards the Sun. 

%%%%%%%%%%
\section{Discussion}

This work is under the assumption that the WIMPs in the solar halo are not disturbed during the capture process. In reality the Solar System also has planets, whose gravitational interaction can disturb the orbiting WIMPs. This effect, as caused by Jupiter, has been discussed in e.g. \cite{Peter}. A massive planet, like Jupiter, can disturb orbiting WIMPs so that they end up on orbits no longer intersecting the Sun, potentially making them very long lived. On the other hand, WIMPs on orbits stretching so far out spend almost all their time far away from the Sun and hence do not contribute so much to the density close to the Sun. Furthermore, planetary interactions are more likely to throw WIMPs out of the Solar System, reducing the WIMP density. WIMPs disturbed to bound orbits not intersecting the Sun end up on orbits sensitive to planetary interactions and will, in time, be disturbed again and typically lose their future contributions to the density close to the Sun. There is, however, a possibility that some fraction of these WIMPs on non-solar-crossing orbits turn out to be more stable, rendering a possible source for higher WIMP densities in the Solar System. Such a mechanism has been discussed in \cite{dk1,dk2}, where it was argued that WIMPs which scatter in the outskirts of the Sun can be perturbed by the planets to orbits not crossing the Sun, in such a way that the new orbits become stable over very long timescales. This WIMP population would then build up over time and it was argued in \cite{dk1,dk2} to be a significant source of WIMPs in the Solar System. These calculations were analytical, but they have not been confirmed in later numerical simulations \cite{ap1}. In fact, in \cite{ap1}, it is argued that the life-times would be much shorter and that the density enhancement would be quite small. In principle, it would be interesting to add planetary interactions to our simulations, as these could in principle increase the density somewhat (even if not as much as originally argued in \cite{dk1,dk2}). That is however outside the scope of this paper, as it would be a very extensive task to do. It is also unlikely that the effects would be large enough to be interesting.

Another possibility to get higher WIMP densities is from a dark matter disk \cite{Read:2009iv}, which could enhance the capture rate in the Sun by up to an order of magnitude \cite{Bruch:2009rp}. This is predominantly because WIMPs from the dark disk pass at lower velocities and are hence easier for the Sun to capture as the energy loss required in the scatter is lower. However, lower initial velocities also imply that the just captured WIMPs will typically have lower energies, altering the distribution in Fig.~\ref{fig:scatter}. This implies that the captured WIMPs will survive less scatters before being fully trapped inside the Sun, which could reduce the density contribution per captured WIMP. Hence, the dark disk scenario need not increase the density around the Sun as much as it increases the capture rate. Even under the optimistic assumption that the density is effected as much as the capture rate, which would increase the gamma-ray signal by up to  two orders of magnitude, the signal is still too low to be detectable.

The calculated gamma-ray flux at Earth from WIMP annihilations around the Sun is extremely low, as seen in table \ref{table:flux}. The calculated flux, at best, corresponds to around one gamma photon per (10 km)$^2$ per century, which is hardly ever detectable. 

The low calculated signal in this work is in contradiction with the earlier result published by Strausz \cite{Strausz}. The analysis conducted here is, however, far more detailed. In \cite{Strausz} it is assumed that the WIMPs move in one-dimensional orbits and that they all have some sort of average properties, such as the average energy loss in a scatter and the average scatter probability in a passage through the Sun. It is, however, very unlikely for the assumptions in \cite{Strausz} to, by themselves, be responsible for such a strong disagreement. The passage where most of the disagreements appear is not explained in detail in \cite{Strausz}, and hence, it is impossible to tell what the source is of the difference in results. 

One should note also, that a calculation of the WIMP densities in the halo has also been performed by Fleysher \cite{Fleysher:2003iya}. Fleysher finds even larger densities than Strausz, but that calculation cannot be correct. The solar halo WIMP densities in the calculation by Fleysher are proportional to the scattering cross sections, which is unreasonable, as the scattering cross sections that enter in the probability of the first scatter and the lifetime of the WIMP orbits cancel (see section \ref{sec:sigmadep} for a more thorough discussion about this).

The conclusion of the work presented here, on the other hand, agrees with the conclusion argued by Hooper \cite{Hooper}. That work is also not very detailed but more recent than Strausz' paper. Also Peter \cite{Peter} has performed simulations of WIMPs in the Solar System, from which the WIMP density can be extracted. As discussed in \cite{Peter:2009qj} those results are very similar to the results found here. Another test of the validity of our results is performed in Appendix \ref{sec:reasonable}.

Finally, we can note that we set out this work with the hope of proving Strausz' optimistic calculations to be correct, but unfortunately, our detailed calculations instead show that the fluxes are indeed far too small to be detectable. On top of that, there are astrophysical backgrounds in which the signal is very well hidden. To reach this conclusion we have made optimistic assumptions about the WIMP dark matter candidate. However, we have assumed that the WIMPs are thermally produced in the early Universe, i.e.\ has an annihilation cross section of the order of $3\times 10^{-26}$ cm$^{3}$ s$^{-1}$. Of course, non-thermal productions or other enhancements (like a Sommerfeld enhancement at low annihilation velocities) would also be possible, but given the large enhancements needed to get an observable flux of gamma-rays from WIMP annihilations around the Sun, there are far better ways to search for those WIMPs in those cases, like gamma-rays from dwarf galaxies, annihilations in the Galactic halo, or effects on the cosmic microwave background radiation \cite{Galli:2009zc}. 

\appendix
\section{Are the results reasonable?}
\label{sec:reasonable}
The calculated signal is truly very weak. This appendix is a further check if the results are reasonable by also estimating the density using simpler arguments. Using what we have learned about the capture process and the shape of the WIMP density profile one can, assuming that these qualitative observations are valid, quite easily make an upper estimate of the magnitude of the WIMP density. This section studies the WIMP configuration $M=100$ GeV, $\sigma_\mathrm{SD}=10^{-3}$ pb and $\sigma_\mathrm{SI}=0$.

We can compare our total capture rate of WIMPs in the Sun with that estimated from previous work \cite{darksusy}. For the WIMP configuration chosen the WIMP capture rate is then $\Gamma_\mathrm c=2.4\times 10^{23}\textrm{ s}^{-1}$. Using this result, an upper estimate of the WIMP population available to build up the WIMP halo can be found by estimating an upper limit of the typical time scale for the WIMP capture process. %{\sf DarkSUSY}

From the Monte Carlo simulations one finds it to be extremely unlikely for a WIMP of this kind to need more than 50 scatters for complete solar entrapment, which is easily verified to be reasonable by looking at the kinematics of the scatters. We assume that our hypothetical WIMP will lose the same amount of energy in each of its 50 scatters, which gives an upper estimate of the entrapment time because the assumption boosts the proportion of large orbits.

The next assumption is that our WIMP has zero total energy before its first scatter. By this assumption the largest bound orbits are not taken into account, however, these orbits do not contribute much to the WIMP density close to the Sun, since a particle in this kind of orbit spends most of its time in the outer regions of the orbit and moves at maximum speed in the inner part.

The reduced energy of the WIMP needs to be less than $\sim -10^{11}$ m$^2$s$^{-2}$ for the orbit to be completely trapped inside the Sun. This should here happen at scatter number 50; the WIMP's reduced energy after scatter number $k$ is then $\mathcal E(k)=-2\times 10^9\mbox{ }k \mbox{ \ m$^2$s$^{-2}$}$. The WIMP's energy after its first scatter then gives an orbit stretching out to $r_{\max}=95R_\odot$, which is more exact than the estimate above since the orbit needs a comparatively very low angular momentum to still cross the Sun, and $r_{\max}$ can then be determined quite well.

From these assumptions one gets an upper estimate of the total time the entrapment process takes a WIMP
\begin{equation}
T_{\mathrm{tot}}=\frac{1}{P}\frac{\pi GM_\odot}{\sqrt 2}\sum_{k=1}^{50} \frac{1}{(-\mathcal E(k))^{3/2}}=\frac{1}{P}\frac{\pi GM_\odot}{4\times 10^{27/2}}\sum_{k=1}^{50}\frac{1}{k^{3/2}} = \frac{7.7\times 10^6}{P} \mbox{ \ s}, \label{total_time}
\end{equation}
where $P$ is a lower estimate of the average scatter probability per solar passage for a WIMP in a bound orbit. The time required for a revolution in a given orbit is given by Eq.~(\ref{orbit_time}). It might be worth noting that this total time for the capture process also includes the time the WIMP spends inside the Sun.

The total number of WIMPs orbiting the Sun at any given moment is then $N=\Gamma_\mathrm c T_{\textrm{tot}}$ and as discussed above they maximally stretch out to $95R_\odot$. Assuming that the shape of the WIMP density curve in Fig.~\ref{fig:finalsdsi}a is correct, the WIMP line density is proportional to $r^{-0.48}$. This gives the WIMP number density 
\begin{equation}
n=\frac{0.52N}{4\pi (95^{0.52}-1)R_\odot^{0.52}} r^{-2.48}=\frac{2.0\times 10^{23}}{P}r^{-2.48} \mbox{ \ m$^{-3}$}
\end{equation}
and hence, the gamma-ray flux at Earth, cf. (\ref{total_flux}) 
\begin{equation}
\Phi_\gamma =\frac{7.7\times 10^{-25}}{P^2}\mbox{ \ m$^{-2}$ s$^{-1}$}.
\end{equation}

What is left to estimate is the average scatter probability per solar passage. In the Monte Carlo simulations it was found for our WIMP scenario that the vast majority of the orbits has a scatter probability between $10^{-3}$ and $10^{-4}$ per solar passage, orbits with scatter probabilities of less than $10^{-6}$ are found to be very rare. In this approximation we set $P=10^{-4}$. This corresponds to the scatter probability for a WIMP that passes through the Sun on a straight line with minimum distance to the solar centre of 40\% of the solar radius, which makes a reasonable lower estimate since most orbits come closer than that to the solar centre. It is easily verified that this probability is reasonable for the column density in the Sun.

For this WIMP configuration the upper estimated flux of gamma-rays more energetic than 1~GeV is at Earth then finally
\begin{equation}
\Phi_\gamma =8\times 10^{-17} \mbox{ m}^{-2}\mbox{s}^{-1},
\end{equation}
which is as expected a few orders of magnitude higher than the calculated flux from the Monte Carlo simulations of $8.4\times 10^{-19}\textrm{ m}^{-2}\textrm{s}^{-1}$, cf. table \ref{table:flux}. The upper limit on the gamma-ray flux is too low to be detected. This reasoning is valid as long as a few long-lived WIMPs do not dominate the WIMP density.

%%%%%
\begin{acknowledgments}
We thank the Swedish Research Council (VR) for support. We also thank Tommy Ohlsson for pushing us to publish this paper.

%This work was supported by the Swedish Research Council
%(Vetenskapsr\aa det), contract no. 315-2004-6519 (S.S.).
\end{acknowledgments}
\bibliographystyle{h-physrev5-je}
\bibliography{wimphalo}

\end{document}